\documentclass[useAMS,usenatbib]{mn2e}

\title[Centrally condensed turbulent cores: Massive stars or fragmentation?]
{Centrally condensed turbulent cores: Massive stars or fragmentation?}
\author[Clare L. Dobbs, Ian A. Bonnell and Paul C. Clark]{Clare L. Dobbs,\thanks{E-mail:
cld2@st-and.ac.uk} 
Ian A. Bonnell and Paul C. Clark\\
School of Physics and Astronomy, University of St Andrews, 
North Haugh, St Andrews, Fife, KY16 9SS}
\usepackage{amssymb}
\usepackage{amsmath}
\usepackage{psfig}

\begin{document}

\date{Accepted}

\pagerange{\pageref{firstpage}--\pageref{lastpage}} \pubyear{2004}

\maketitle

\label{firstpage}

\begin{abstract}
We present numerical investigations into the formation of massive stars from
turbulent cores of density gradient $\rho \propto r^{-1.5}$.
The results of five hydrodynamical simulations are described, following the
collapse of the core,
fragmentation and the formation of small clusters of protostars.
We generate two different initial turbulent velocity fields corresponding to
power-law spectra $P \propto k^{-4}$ and $P \propto k^{-3.5}$, and apply two
different initial core radii. Calculations are included for both completely
isothermal collapse, and a non-isothermal equation of state 
above a critical density ($10^{-14}$gcm$^{-3}$).
Our calculations reveal the preference of fragmentation over monolithic star
formation in turbulent cores. Fragmentation was prevalent in all the
isothermal cases. Although disc fragmentation  was largely suppressed
in the non-isothermal runs due to the small dynamic range between the initial density
and the critical density, our results show that some
fragmentation still persisted. This is inconsistent with
previous suggestions that turbulent cores result in the formation of a 
single massive star. We conclude that turbulence cannot be measured as an isotropic pressure
term. 
\end{abstract}


\section{Introduction}

There are several potential difficulties in forming high-mass stars. Firstly, the
timescale of less than $10^6$ years to assemble 10 to more than 100 $M_{\odot}$  implies
large accretion rates \citep{Zin1993,BM2001,Norberg2000}. Secondly, their crowded location 
in the centre of young stellar clusters \citep{CBH2000,HillH1998,Carp1997,Lada2003} 
limits the final mass of any collapsing fragment:   
confining the Jeans radius to be less than the interstellar separation 
means that the
resultant Jeans mass will be correspondingly small \citep*{Zin1993,BBZ1998}.
Lastly, and most importantly, the radiation pressure from a high-mass star is sufficient
to reverse the infall of gas that contains typical dust properties \citep{
Wolfire1987,Beech1994}.
There are a number of ways this last problem can be circumvented. The radiation
pressure could be overwhelmed by ultra-high accretion rates \citep{McKee2003}.
Alternatively, accretion could
occur preferentially through an equatorial disc, while the central object is rapidly
rotating and thus emitting most of its radiation towards the poles \citep{Yorke2002}. 
A third solution is that massive stars form due to stellar mergers in the
ultra-dense core of a cluster \citep{BBZ1998, BB2002}.

The \citet{McKee2003,McKee2002} model depends on high accretion rates that
would be expected to occur in a dense gas core in the centre of a cluster. This 
core is envisioned to
be supported by turbulence, as thermal pressure is inadequate, and adopts a 
steep density profile in order to prevent fragmentation.
Even neglecting how such a core could arise in the centre of a stellar cluster, one
potential difficulty is that turbulent clouds are known for their tendency to fragment
and form a stellar cluster.  It is this possibility that we address here.

Fragmentation during the initial stages of star formation
has been illustrated through numerical studies of collapsing turbulent molecular
clouds. \citet*{BBB2003} describe the collapse of a uniform $50 M_{\odot}$ 
turbulent cloud of diameter 0.375pc, which forms 3 cores containing 23+ protostars
and 27+ brown dwarfs. Fragmentation has also been demonstrated in isolated,
rotating $1 M_{\odot}$ cloud cores \citep*{BuBB1997}, where low
multiple systems are produced despite an initial $\rho \propto r^{-1}$ density
distribution.  

The formation of stellar clusters from turbulent cloud cores \citep*{BBV2003}
has shown how the fragmentation of the turbulently induced filamentary structure produces hundreds of stars. These stars  fall into local potential
wells forming small sub-clusters which eventually  merge to form a large stellar cluster.  The massive stars form in the
centre of the sub-clusters due to competitive accretion, naturally explaining why
massive stars are found in the cores of dense stellar clusters \citep*{BBV2004}. The stellar densities
in these cores can be very high, approaching that for stellar mergers to occur 
\citep{BB2002}. 

Our calculations adopt a 
turbulent core with a density profile and velocity dispersion
comparable to \citet{McKee2002}. The purpose of this paper is to investigate 
whether the steep density gradient assumed by 
McKee and Tan is sufficient to prevent fragmentation and consequently provide a suitable 
approach to massive star formation.

\section{Computational Method}

\subsection{SPH code}

We use the 3D smoothed particle hydrodynamics (SPH) code based on the version by
Benz \citep{Benz1990}. The smoothing lengths between particles are allowed to vary,
but the typical number of neighbours for each particle is $N_{neigh} \thicksim
50$.  Artificial viscosity is included with the standard parameters $\alpha=1$
and $\beta=2$. Gravitational forces are calculated using a binary tree. The code uses 
$10^6$ particles (so for a $30M_{\odot}$ core (Section~2.3) the minimum particle mass is
$3 \times 10^{-5} M_{\odot}$) and simulations were run for approximately half a 
free-fall time. All computations were performed using the United Kingdom's
Astrophysical Fluids Facility (UKAFF), a 128 CPU SGI Origin 3000 supercomputer.

\subsection{Sink particles}
Simulations in SPH become very computationally expensive as the density
increases during the collapse of a core. The insertion of sink particles 
\citep*{Price1995}
at a certain density is widely used to extend 
calculations.
To resolve the local Jeans mass requires $\thicksim 2N_{neigh}$ particles 
\citep{BBu1997}, i.e. $3 \times 10^{-3} M_{\odot}$ in these simulations. 
This is then the minimum resolvable mass that can be considered unstable for collapse.
The minimum density for insertion of a sink particle can then be determined from 
Jean's equation, inserting $3 \times 10^{-3} M_{\odot}$ as the
Jean's mass:   
   
\begin{equation}
M_J=\bigg(\frac{5 R_g T}{2 \mu G} \bigg)^{3/2} \bigg(\frac{4}{3} \pi \rho
 \bigg)^{-1/2}=3 \times 10^{-3} M_{\odot}
\end{equation}
Taking the temperature $T$ to be $20K$ and $\mu=2.46$, this rearranges to 
give $\rho_{acc} = 1.06 \times 10^{-13}$gcm$^{-3}$. The corresponding 
radius for accretion is $r_{acc} \thicksim 20$AU.

\subsection{Initial conditions}
McKee \& Tan (hereafter MT(2003)) assume a centrally condensed core, following a 
density profile of 
$\rho \propto r^{-k_{\rho}}$. We take $k_{\rho}=1.5$, the fiducial value of
MT(2003). This is 
shallower than the singular isothermal sphere model ($k_{\rho}=2$, \citep{Shu1977})
but comparable with observations of 
cloud cores (\citet{MF1992} give $k_{\rho}=1.6$, \citet{WT1994}
$k_{\rho}=1.2$). 

The first computations took the core radius $R$ to be 0.06pc with a mass of $30M_\odot$.
With a resolution of $10^6$ particles, the initial
density profile $\rho \propto r^{-3/2}$ rendered a central density of 
$\thicksim 3 \times 10^{-15}$gcm$^{-3}$, whilst the average density was 
$\rho_{av}\thicksim 2.5 \times 10^{-18}$gcm$^{-3}$. 
The results are given in terms of the free fall time, $t_{ff}=(3 \pi / 32 G \rho_{av}) 
\thicksim 4.5 \times 10^4$ years. 
MT(2003) use similar initial conditions, but take a mass of $60M_{\odot}$
to form a $30M_\odot$ star with 50\% efficiency.
We also performed computations with a core radius 0.2pc, since this gave similar 
dimensions to \citet{BBB2003}. The average density was then $3 \times 10^{-19}$gcm$^{-3}$ 
and the central density $2 \times 10^{-16}$gcm$^{-3}$.
The free fall time for these results was
$2.7 \times 10^5$ years. In all calculations, the initial temperature of the core was 20K, comparable
to the cold gas of which the first massive stars form. 

As in equation (1), the Jeans mass is
\begin{equation}
M_J (\rho(r))=\bigg(\frac{5 R_g T}{2 \mu G} \bigg)^{3/2} \bigg(\frac{4}{3} \pi \rho
 \bigg)^{-1/2}
\end{equation} 
where $\mu$ the mean molecular weight is taken as 2.46. The number of Jeans masses contained
in a core is then
\begin{equation}
N_J (R)=\int^{R}_{0} \frac{\rho(r) 4 \pi r^2}{M_J(\rho(r))} \thickspace dr.
\end{equation}
This gives 69 Jean's masses when $R=0.06$pc and 11.5 when $R=0.2$pc.

\subsubsection{Turbulence}
MT(2003) incorporate turbulence by means of an effective turbulent pressure term 
whereby $P\propto~r^{-k_P}$. By assuming hydrostatic equilibrium,
$k_P=2(k_{\rho}-1)$, and the resulting velocity dispersion relation is derived as
\begin{equation}
\sigma \propto c_s = \sqrt{\frac{P}{\rho}} \propto r^{(2-k_{\rho})/2}
\end{equation}
where $\sigma$ is the velocity dispersion, and $c_s$ the effective isothermal sound speed.
Inserting $k_{\rho}=1.5$ then gives $\sigma \propto r^{0.25}$. 
This is shallower than observational results obtained for the similar Larson relation
$\sigma \propto L^{\alpha}$ (L a typical length scale), which indicate that 
$0.25<\alpha<0.75$, e.g. $\alpha=0.38$ \citep{Larson1981}, $\alpha=0.5$ \citep{Myers1983}.

We simulate turbulence using a grid method outlined by \citet{Dub1995} and briefly described
here. The initial velocity field is calculated from a random Gaussian
distribution and follows a power spectrum
\begin{equation}
P(k) \equiv \thickspace <|v_k|>^2 \thickspace \propto k^{-n},
\end{equation}
where $k$ is the wavenumber. To obtain a divergence free velocity field, 
velocities are constructed from a vector potential $\mathbf{A}$, such that
$\mathbf{v}=\nabla \times \mathbf{A}$. We sample components of $\mathbf{A_k}$, the Fourier
transform of  $\mathbf{A}$, from a grid of co-ordinates $(k_x,k_y,k_z)$.
In order to generate a Gaussian distribution satisfying (5), components of 
$\mathbf{A_k}$ are selected as 
\begin{equation}
\mathbf{A_{k}}(k_x,k_y,k_z)= |\mathbf{k}|^{\frac{-n-2}{2}} (C_{k_x}  
e^{i \theta_{k_x}}, C_{k_y} e^{i \theta_{k_y}},
C_{k_z} e^{i \theta_{k_z}})
\end{equation}
The amplitudes $C$ are sampled from a Rayleigh
distribution and phases $0< \theta < 2\pi$ from a random distribution (see \citet{Dub1995}). 
Taking the inverse
Fourier transform of $\nabla \times \mathbf{A_k}$ then gives
\begin{equation}
\mathbf{v}=\frac{1}{2 \pi} \sum_{\mathbf{k}} i \mathbf{k} \times \mathbf{A_k}
\thickspace e^{i \mathbf{k}.\mathbf{r}} 
\end{equation}
Numerically the real $x$ component of (7) becomes 
\begin{equation}
\begin{split}
v_x = & \frac{1}{2 \pi} \sum^{k_{max}}_{-k{max}}  
|\mathbf{k}|^{\frac{-n-2}{2}} 
\thickspace  \Big[ k_z C_{k_y} \sin(\mathbf{k} .\mathbf{r}+\phi_{k_y}) \\ 
& -  k_y C_{k_z} \sin(\mathbf{k} . \mathbf{r}+\phi_{k_z}) \Big] 
\end{split}
\end{equation}
summed over $k_x,k_y$ and $k_z$. Similar expressions provide $v_y$ and $v_z$, for a
a particle at position (x,y,z).
The parameter $k_{max}$ in equation (8) is often chosen to satisfy the resolution 
requirements of the model, 
i.e $k_{max} \thicksim (2 \pi / L_{min})$ where $L_{min}$ is the smallest length scale or minimum separation
between particles. However, since our model is centrally condensed and the density varies considerably,
we allow ${k_{max}}$ to vary for each particle according to the local smoothing length. We take
\begin{equation}
(k_{max})_i=\text{INT} \bigg(\frac{2 \pi} {\thickspace h_i}\bigg)
\end{equation}
$h_i$ is the smoothing length of the $i$th particle.
In this way, velocities in less dense (low resolution) regions, 
could be computed much faster than those in the dense centre of the core (high resolution),
reducing the computational expense of the calculation. 
  
The velocity field generated gives a dispersion - 
length scale dependence similar to the Larson relation. 
Calculations are included for $\sigma \propto r^{0.25}$, consistent
with MT(2003), and $\sigma \propto r^{0.5}$ which corresponds better with observations.
These dispersion laws are attained by generating power spectra with 
$n=3.5$ and $n=4$ respectively. However \citet{Myers1999} show that $n\thicksim3.25$ 
provides a better 
fit for the $\sigma \propto r^{0.25}$ dispersion law.

The core is initially in virial equilibrium, such that
\begin{equation}
E_{grav}+2E_{turb}+2E_{therm}=0. 
\end{equation} 
The ratios of turbulent to gravitational energy were
\begin{eqnarray}
\frac{E_{turb}}{E_{grav}} & \thicksim 0.4 & \text{for} \quad R=0.06pc, \\
\frac{E_{turb}}{E_{grav}} & \thicksim 0.25 & \text{for} \quad R=0.2pc.
\end{eqnarray}
The turbulent energy is then allowed to dissipate over the
dynamical timescale of the core \citep{Mac1997}. We do not include any driving, which if
included, should increase the susceptibility to fragmentation.

\subsubsection{Equation of State}
The thermal behaviour of the cloud core was assigned as either:
\begin{enumerate} 
\item Completely isothermal.
\item Non-isothermal above a critical density, allowing the 
gas to heat up as the core collapses.
In this case the equation of state exponent varies as  
\begin{equation}
 \gamma=\left\{\begin{array}{cc}
 1 & \textrm{if $\rho < 10^{-14}$gcm$^{-3}$} \\ 
 1.67 & \textrm{if $\rho \geq 10^{-14}$gcm$^{-3}$}
 \end{array} \right.
\end{equation}
\item Non-isothermal above the critical density with
a second isothermal collapse phase for $\rho > 10^{-12}$gcm$^{-3}$.
\end{enumerate}
Although the core is initially globally in virial equilibrium, the inner
regions are actually over-supported due to the isothermal nature of the
gas.

\section{Results}

A summary of the calculations undertaken, their parameters and the overall results
are shown in Table~1. The parameter $\alpha$ describes the initial velocity dispersion
(where $\sigma \propto r^{\alpha}$).

In all simulations, turbulence initially supports the core in virial 
equilibrium. As the simulation progresses, turbulence
decays through shocks in the core, dissipating kinetic energy. The core then 
begins to undergo collapse, with density increasing until the formation of individual 
protostars. 
The morphology of the cloud is similar in each simulation - the core
becomes elongated, generating significant structure which forms the basis for subsequent
fragmentation (e.g. Figure~1).
 
\subsection{Isothermal results}
Models 1, 2 and 3 used an isothermal equation of state throughout the simulation. 
Models 1 \& 3 apply a $\sigma \propto L^{0.25}$ dispersion law whereas  
Model~2 uses $\sigma \propto L^{0.5}$ although the initial 
velocity fields are determined by the same series of random numbers in all 
calculations. Model~3 differs by taking a 0.2pc radius cloud. 

All 3 simulations produced significant numbers of protostars. Table~1 includes the total number of
protostars formed, although results cannot be directly compared as some formation was still
ongoing.
 
\begin{table*}
\centering
\begin{tabular}{|c|c|c|c|c|c|c|c|}
\hline
Model & $\alpha$ & R (pc) & Jean  & Predominant time of & No. of & Total mass & Most massive \\
    &          &        &     No.        & formation ($t_{ff}$) & protostars  
    & accreted ($M_{\odot}$) &  protostar ($M_{\odot}$) \\ \hline
1 & 0.25 & 0.06 & 95 & 0.47  & 28 & 2.73 & 0.435 \\ 
2 & 0.5  & 0.06 & 95 & 0.37 & 14 & 2.12 & 0.528 \\ 
3 & 0.25 & 0.2    & 16 & 0.55  & 19 & 1.79 & 0.66 \\ 
4 & 0.25 & 0.06 & 95 & 0.54  & 3$^{\dagger}$  & 2.02 & 0.99 \\ 
5 & 0.25 & 0.2    & 16 & 0.54  & 2  & 2.10 & 1.28 \\ \hline 

\end{tabular}
\caption{Table showing results of all simulations, giving the total number of stars formed and
the approximate time at which most formation occurred. Models 4 \& 5 use a
non-isothermal equation of state above a critical density (Section~2.2.2).
However Model 4~also includes a second isothermal collapse phase.
$^\dagger$Further fragmentation apparent but sink formation suppressed by equation of
state.}  
\end{table*}

\begin{figure*}
\centering
\begin{tabular}{l l l}
\psfig{file=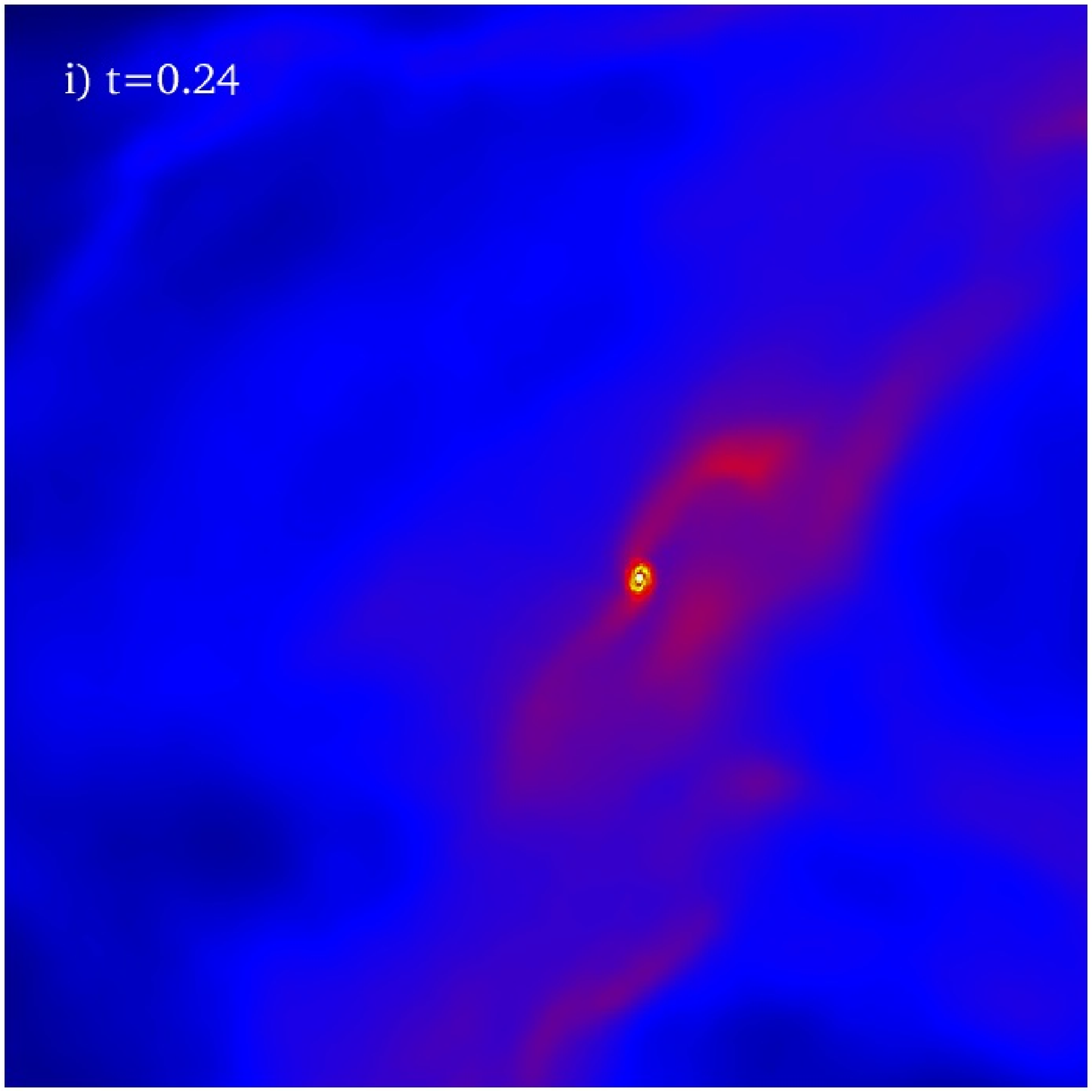,height=2.2in} & &
\psfig{file=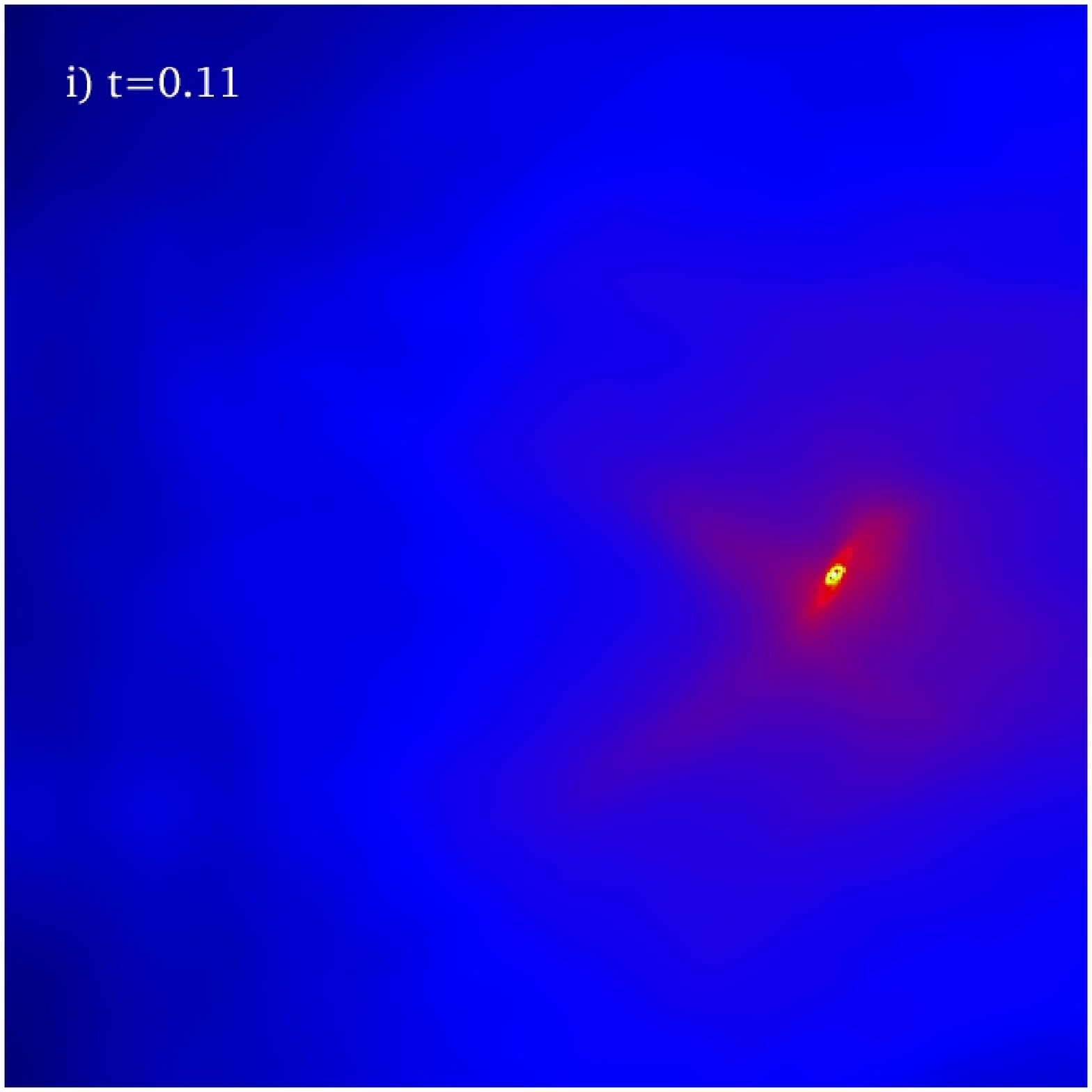,height=2.2in} \\
\psfig{file=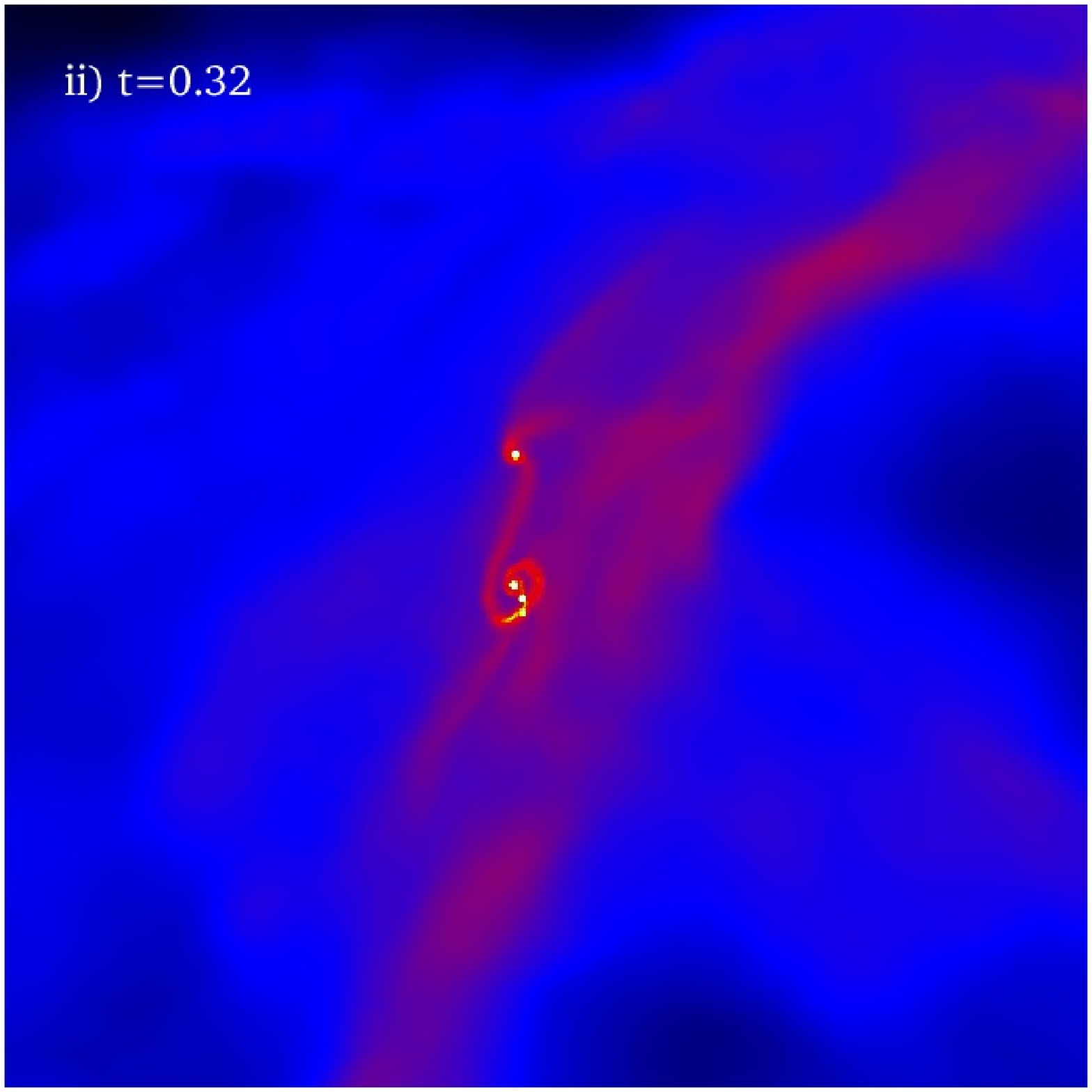,height=2.2in} & &
\psfig{file=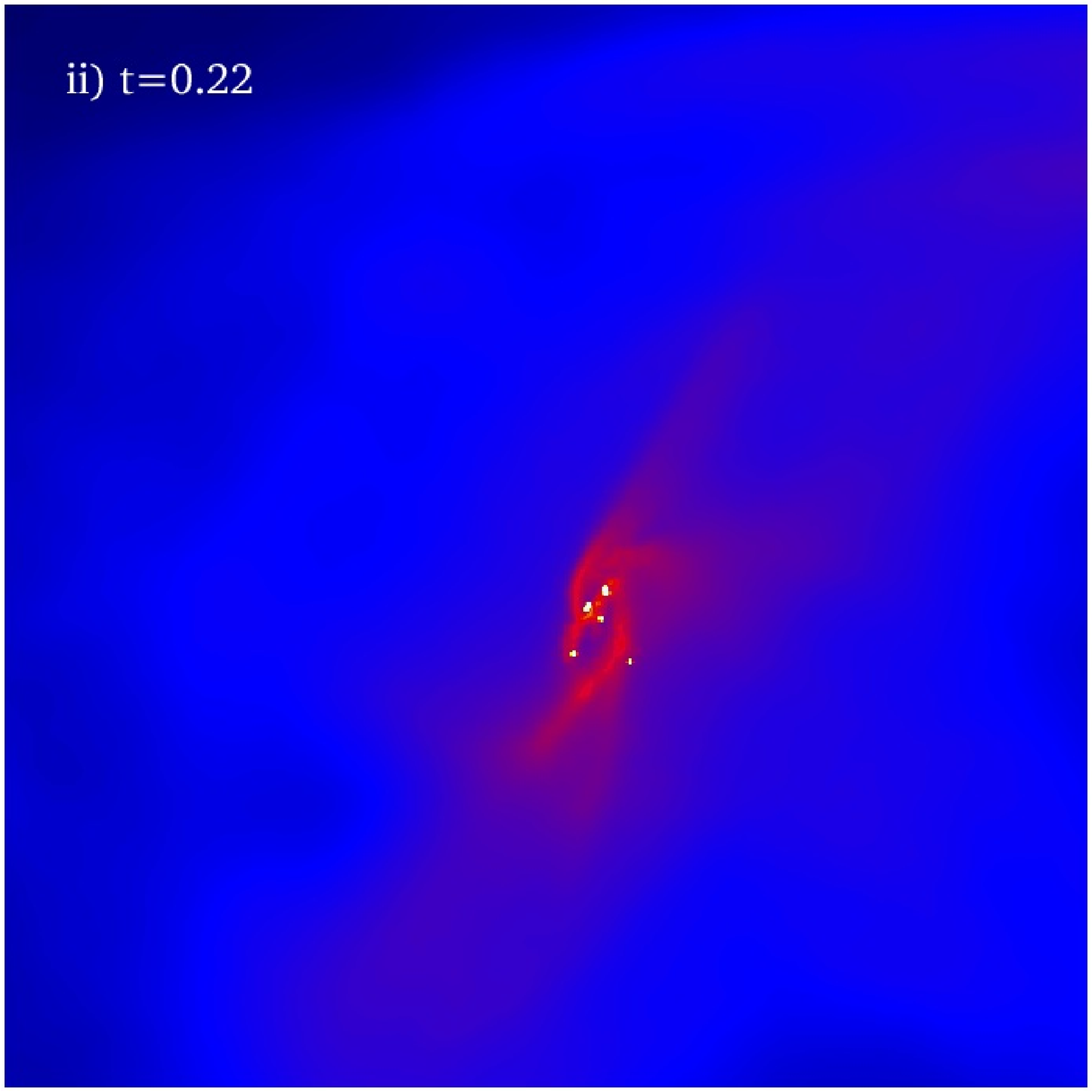,height=2.2in} \\
\psfig{file=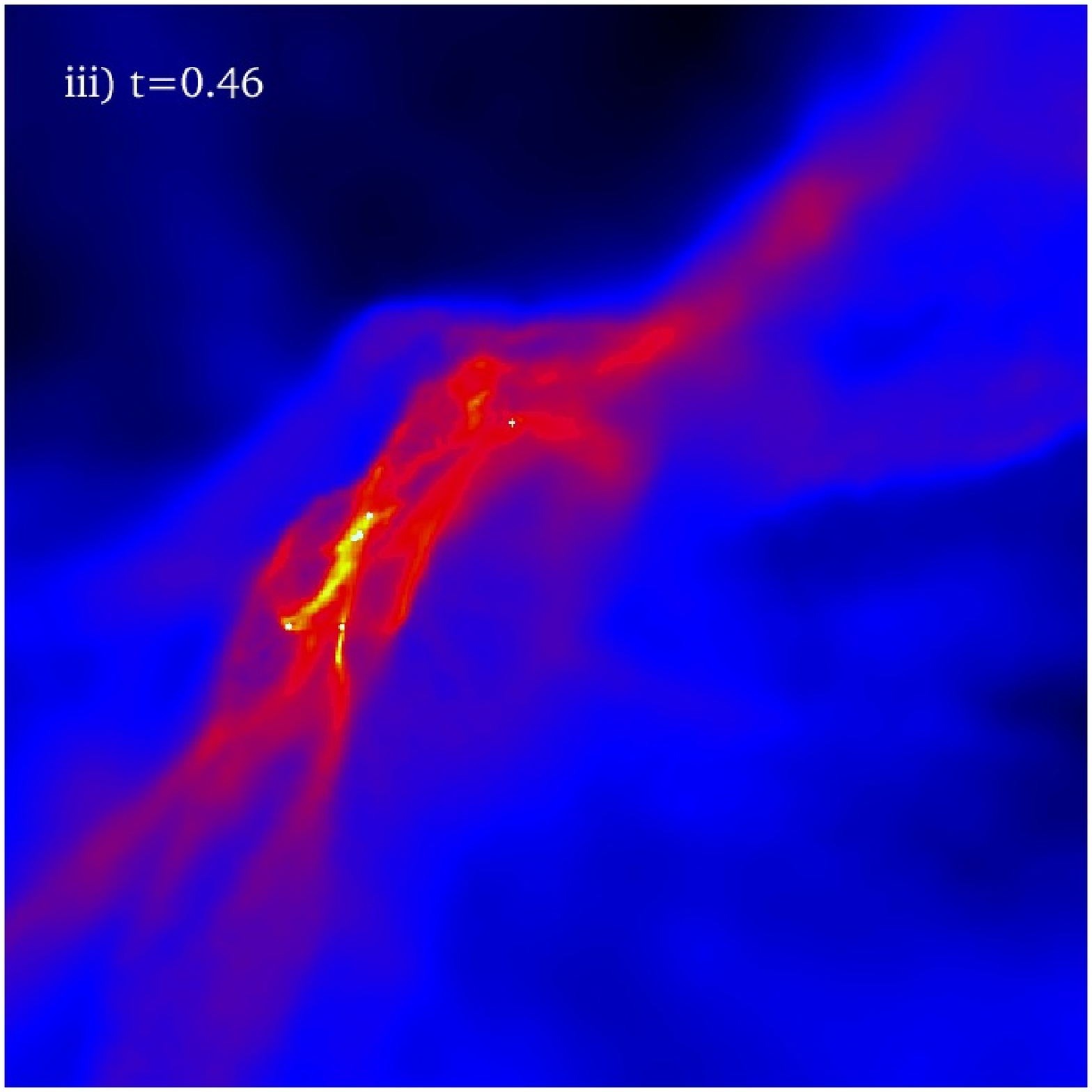,height=2.2in} & &
\psfig{file=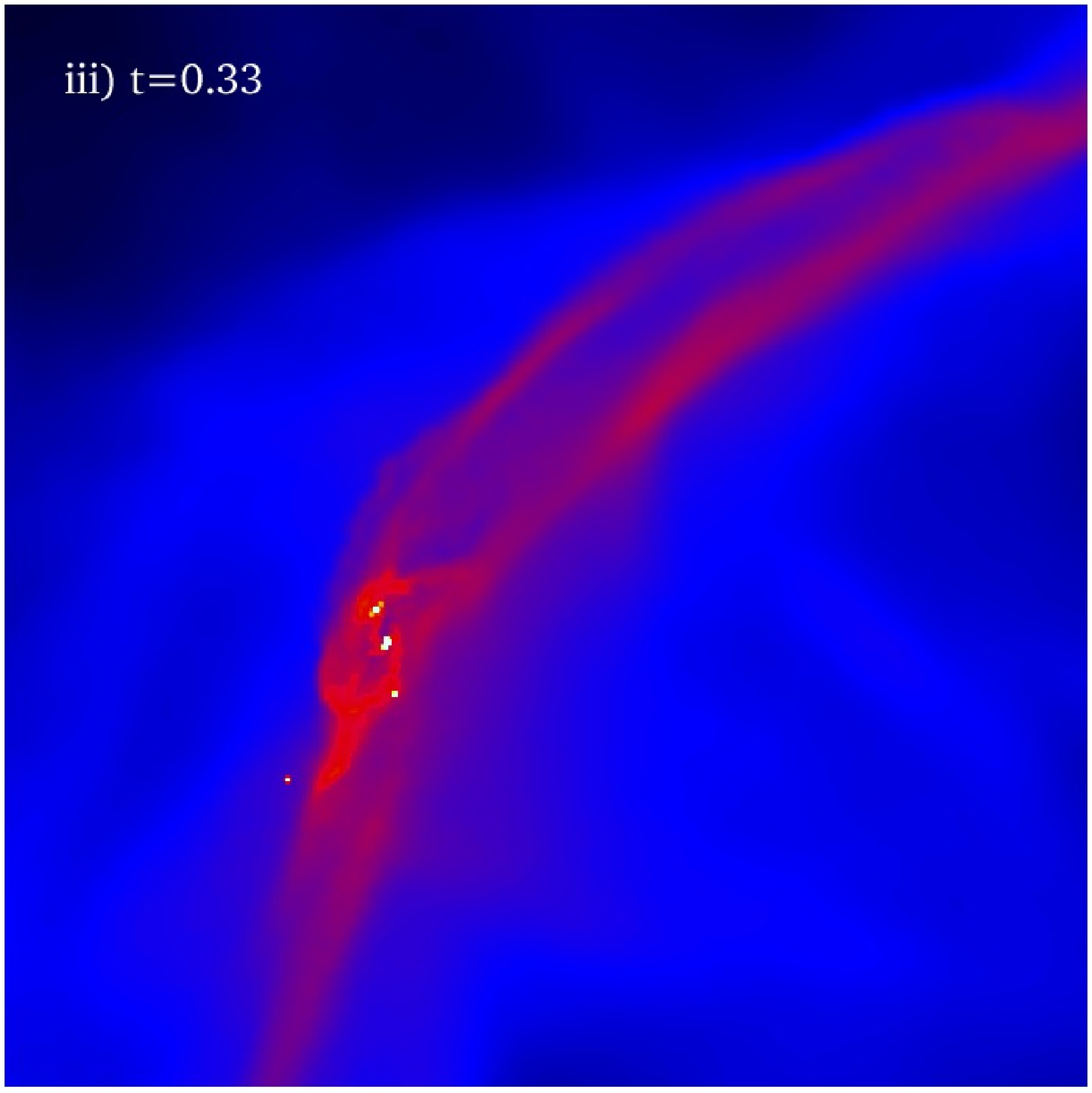,height=2.2in} \\
\psfig{file=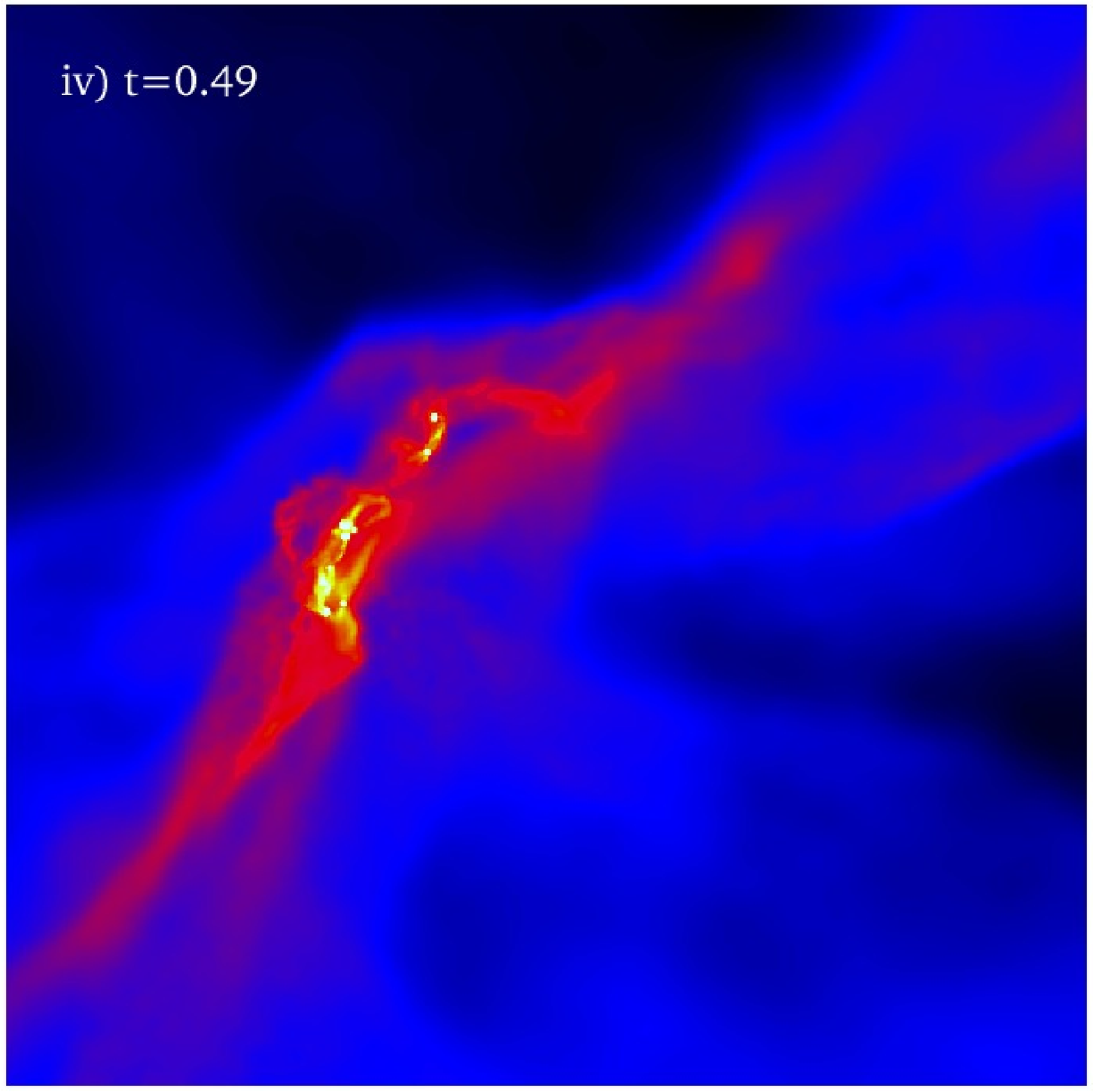,height=2.2in} & &
\psfig{file=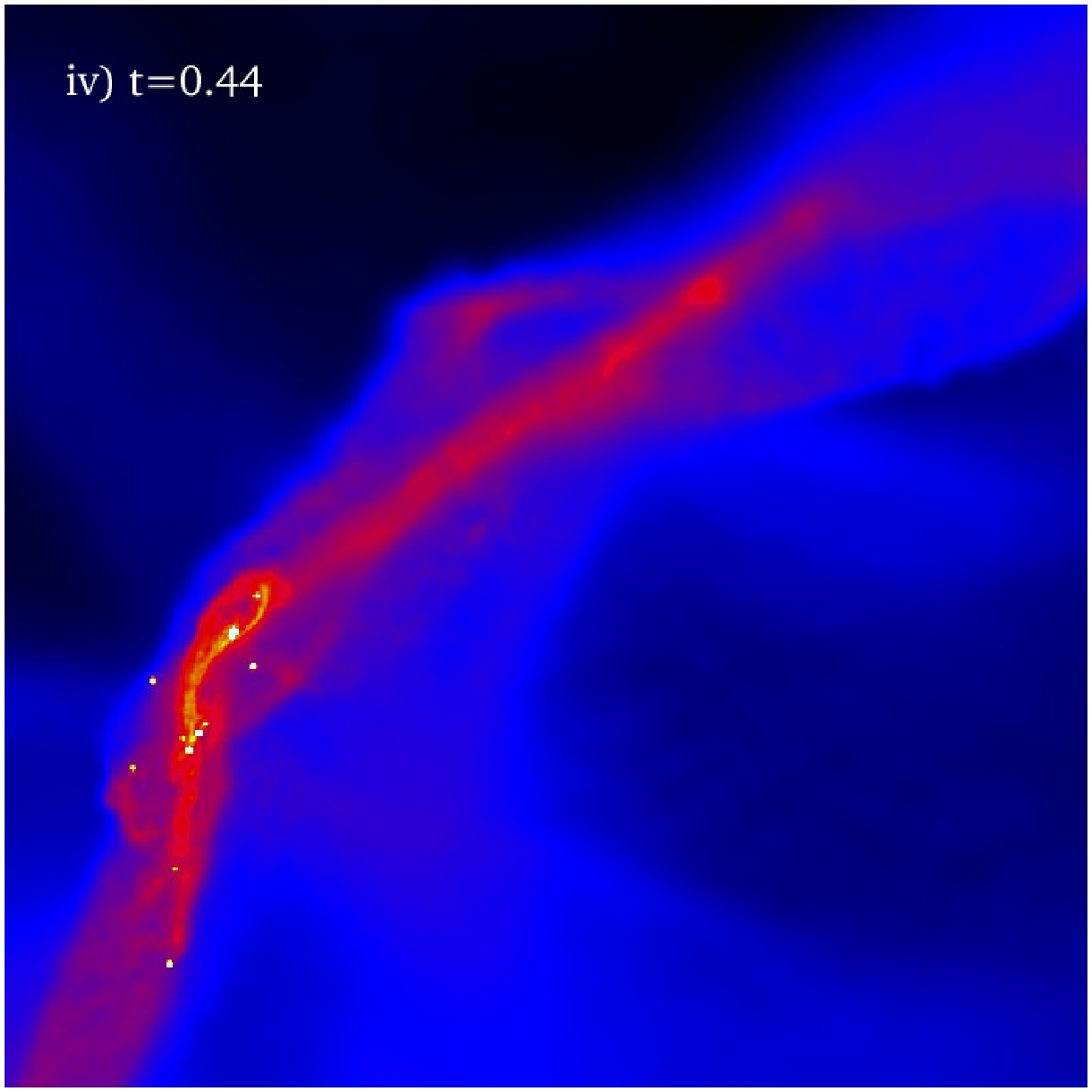,height=2.2in} \\
\end{tabular}
\caption{Column density plots (logarithmic scale, minimum and maximum densities of 
$4 \times 10^{-17}$gcm$^{_3}$ and $4 \times 10^{-13}$gcm$^{_3}$) showing evolution
 of central region of
the core for $\sigma \propto L^{0.25}$ (left) and $\sigma \propto L^{0.5}$ (right). 
Length scale for each plot is 0.02pc
$\times$ 0.02pc. The number of protostars in the plots are: 
Left - i)1, ii)4, iii)6, iv)13; Right - i)1, ii)7, iii)7, iv)14 
. Times are given in units of $t_{ff}=4.5 \times 10^4$ years.}
\end{figure*}

\subsubsection{Comparison between dispersion laws}
Figure~1 displays different stages during Models 1 \& 2, which apply different
power spectra.
The main difference between Models 1 \& 2 is the timescale for protostar
formation and fragmentation. 
Collapse and protostar formation occur later in Model~1 compared to Model~2, although their
overall evolution is similar. This is due to the shallower power law $P \propto k^{-3.5}$
(compared to $P \propto k^{-4}$ for Model 2 when $\alpha=0.5$) which supplies more kinetic
energy over smaller length scales. This provides greater support over small scales and
generates more structure.  
Consequently density plots for Model~1 show more structure than those for Model~2,
and the protostars are less widely distributed (Figure~1).     
 
The left side of Figure~1 illustrates the evolution for the $\alpha=0.25$ dispersion law, which best
represents MT(2003).
A protostar forms in the central part of the core after $0.19t_{ff}$, surrounded by a disc of gas. 
The core fragments after $0.25t_{ff}$, when 2
further protostars form from the disc material. A fourth protostar is also created via fragmentation of an
elongated filament extending from the initial protostar (Figure 1 ii). 
No further protostars form for $\thicksim 0.15t_{ff}$, as the core is largely
supported by internal rotation. However the internal dynamics are still chaotic and further collapse occurs
as support is lost. 
Further fragmentation occurs through collapse along filaments near the core's centre
(Figrure 1 iii). 
This results in a second phase of star formation at around 0.47$t_{ff}$, 
producing a total of 28 protostars after 0.54$t_{ff}$.

Model~2 (Figure~1, right), 
where $\alpha=0.5$, shows similar behaviour. Again an initial protostar is formed after 0.09$t_{ff}$
accompanied by a disc which subsequently distorts and fragments, producing 7
protostars after $0.13t_{ff}$. 
Similarly there is a second phase of fragmentation through the collapse of elongated
filaments, leading to a total of 14 protostars after $0.44t_{ff}$. Figure~2 
plots the mass accreted and the number of protostars
formed against fraction of a free fall time, for both dispersion laws.
Models 1 and 2 display very similar profiles, offset by $\thicksim 0.1t_{ff}$.
The number of protostars formed in each case shows 2 periods of star formation
at 0.1/0.2t$_{ff}$ and 0.38/0.48t$_{ff}$ separated by a period of 0.26/0.15t$_{ff}$. 

\subsubsection{Overall comparison of isothermal runs}
The isothermal results all display a high level of fragmentation, demonstrating that
turbulent pressure is anisotropic. Turbulence cannot be represented by an isotropic thermal
pressure, instead distortion of the core leads to the formation of multiple protostars. 
The degree and timescale of
fragmentation vary according to the turbulent power law
and the size of the core.
Whereas a steeper velocity-sizescale relation ($\alpha=0.25$) advances protostellar 
formation, collapse is delayed for the $R=0.2$pc core in Model~3.   
The larger core has more thermal support so the ratio of thermal to gravitational energy is
greater (equations (11),(12)). 
Protostar formation begins at $\thicksim 0.52t_{ff}$, so collapse of the core and subsequent 
fragmentation occurs at a higher fraction of the free fall time compared with Model~1 where $R=0.06$pc.
Correspondingly, density plots for Model~3 also show less structure, tending 
to retain a
centrally condensed profile, at least over large length scales. There was, however,
still sufficient structure over the centre to produce 19 protostars over $0.6t_{ff}$,
so significant fragmentation still occurred.

\subsubsection{Accretion rates}
The main motivation for MT(2003) was to achieve a high accretion rate
to overcome radiation pressure.
Accretion rates can generally be estimated as the resulting protostellar mass divided
by the free-fall time $(M/t_{ff})$. Thus MT(2003) construct their
models to produce an early accretion rate of $10^{-4} M_{\odot} yr^{-1}$, which 
increases to $10^{-3} M_{\odot} yr^{-1}$.
From Figure~2 we see the accretion rate is initially $\thicksim10^{-4} M_{\odot} yr^{-1}$
increasing to $10^{-3} M_{\odot} yr^{-1}$ after $0.3t_{ff}$.
These accretion rates agree with MT(2003), but are based on the total mass accreted onto
many protostars (between $\thicksim$5-20). The accretion rate for an individual protostar is 
therefore significantly less. In contrast, the cluster accretion models for massive 
star formation \citep{BBV2004}
show that although the mean accretion rate $(M/t_{ff})$ is only $10^{-6} M_{\odot} yr
^{-1}$,
the actual accretion rate onto the growing massive star in the
centre of a cluster is $10^{-4} M_{\odot} yr^{-1}$.

\begin{figure}
\centering
\begin{tabular}{l}
\psfig{file=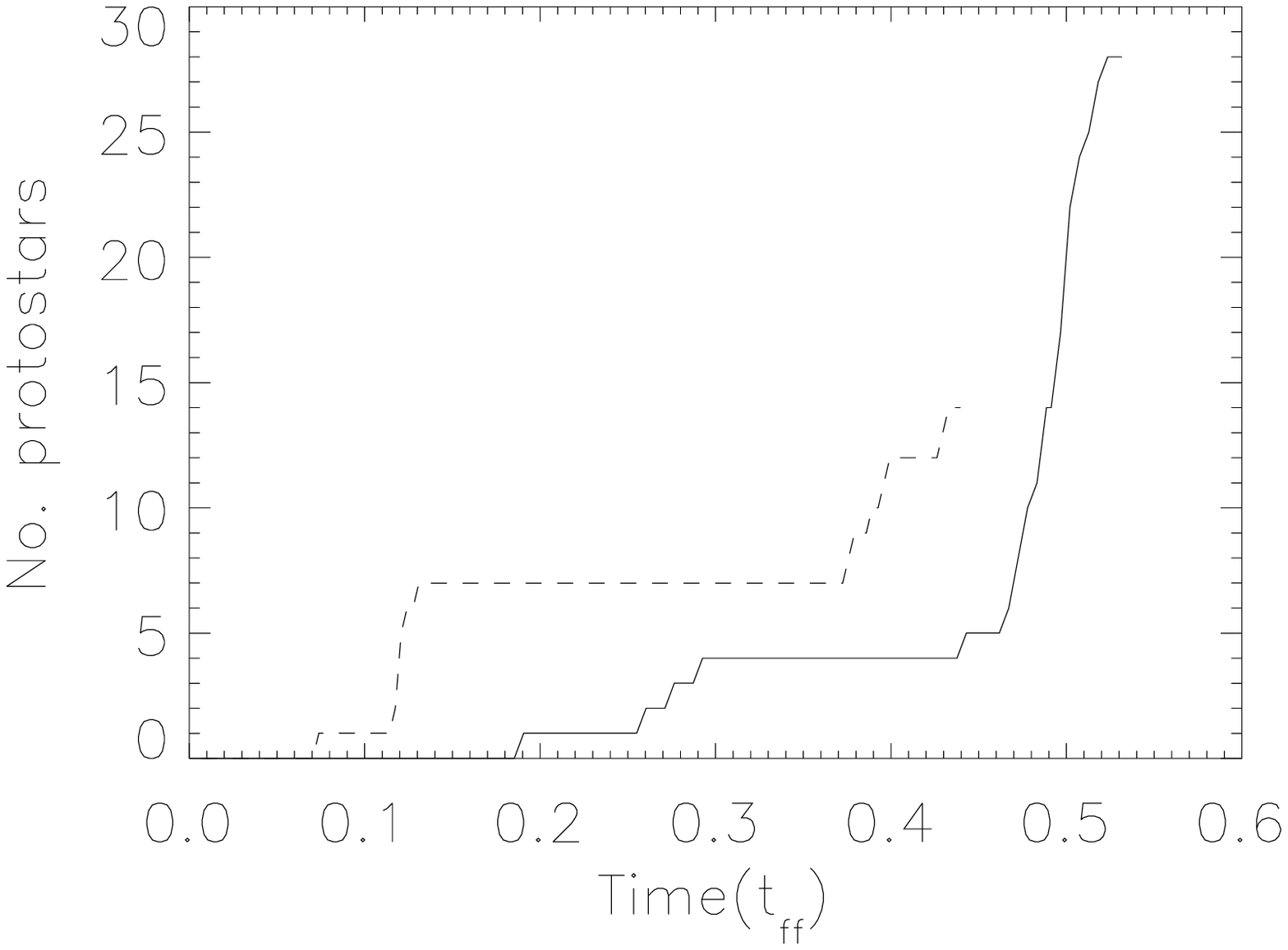,height=2.3in} \\ 
\psfig{file=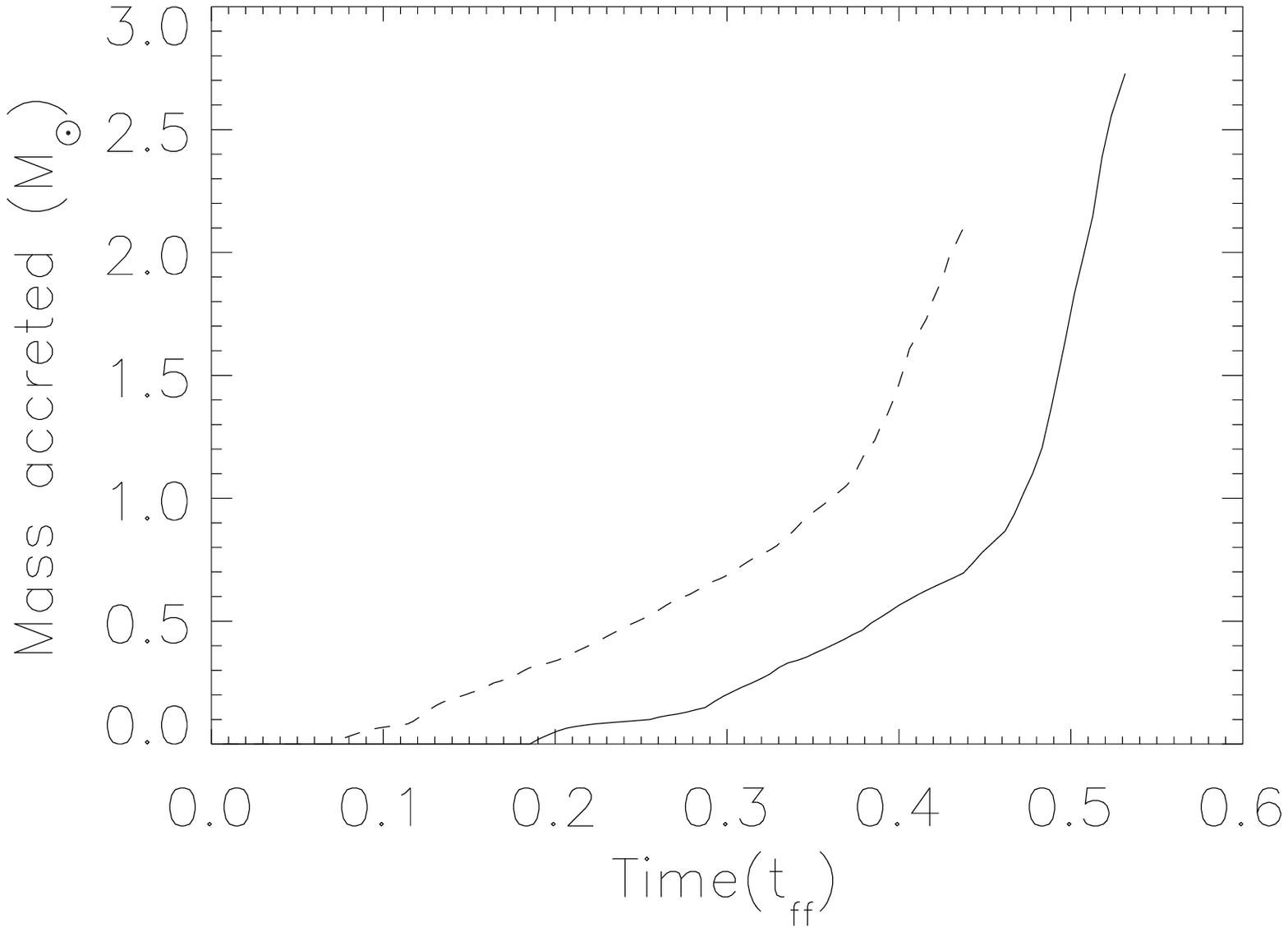,height=2.3in} 
\end{tabular} 
\caption{Plots for number of stars (top) and mass accreted (bottom)
versus time for Model~1,
$\alpha=0.25$ (solid line) and Model~2, $\alpha=0.5$ (dashed line).}
\end{figure}
 
\subsection{Non isothermal equation of state}
The large scale collapse of the non-isothermal models are very similar to the corresponding isothermal
cases, since the equation of state changes only in the central denser regions. However, Table~1 shows 
that the
number of protostars formed is dramatically reduced (Models 4 \& 5). In Model~4, (0.06pc core), 
the equation of state changed before fragments could collapse to form sink particles. 
A second isothermal collapse phase subsequently produced 3 protostars. For Model~5 
(0.2pc core), the
core is initially less dense so 2 protostars were able to form without any second collapse phase.

To understand the difference between protostar formation in the isothermal and non-isothermal models,
it is necessary to compare the protostar-forming region in each case.   
Figure~3 illustrates the contrast between Model~1 (isothermal) and Model~4 (non-isothermal) in the
dense part of the core.
Complex structures continue to exist on small scales in the isothermal case, including a disc and
close companions formed through disc fragmentation. In comparison, Model~4  
is dominated by 2 large volumes of dense thermally supported gas.
 
On scales comparable to the core radius however, density
profiles show that considerable structure still exists in the non-isothermal core.
In Figure~4 (Model~4), the maximum densities over the x and y directions are plotted  
(while the equation of state remains non-isothermal). Five peaks have been selected from the profiles 
which show a large density contrast with the surrounding material. These indicate independent, 
disconnected fragments. Figure~5 shows an accompanying column density plot, obtained at the same time
frame, which
illustrates the large scale structure of the core. The locations of the five peaks are indicated, 
showing where fragmentation is taking place. The 2 highest peaks in the densest part of the core
correspond to the dense regions shown in Figure~3. These also subsequently form 3 protostars during 
the second isothermal collapse phase, as the larger region (of Figure~3) sub-fragments. 
With a longer period of time, and a less severe equation of
state, we would expect all the dense fragments corresponding to these peaks to collapse independently
and form protostars.           

The distortion of the 0.06pc core suggests that monolithic formation is unlikely, despite the
formation of very few protostars. Our analysis suggests fragmentation will occur
producing at least 6 separate bodies (5 peaks identified from Figures~4 \& 5, and isothermal
sub-fragmentation of peak 2). This is again a
consequence of anisotropic turbulence dominating homogeneous thermal support.

Conversely, the larger core (Model~5) remains predominantly centrally condensed. 
Thermal pressure supports the core over large scales, whilst the equation of state
prevents any local fragmentation. With only 2 centrally located protostars, Model~5 would
only be expected to produce a low multiple (2 or 3 protostars) system.     
\begin{figure}
\centering
\begin{tabular}{c}
\psfig{file=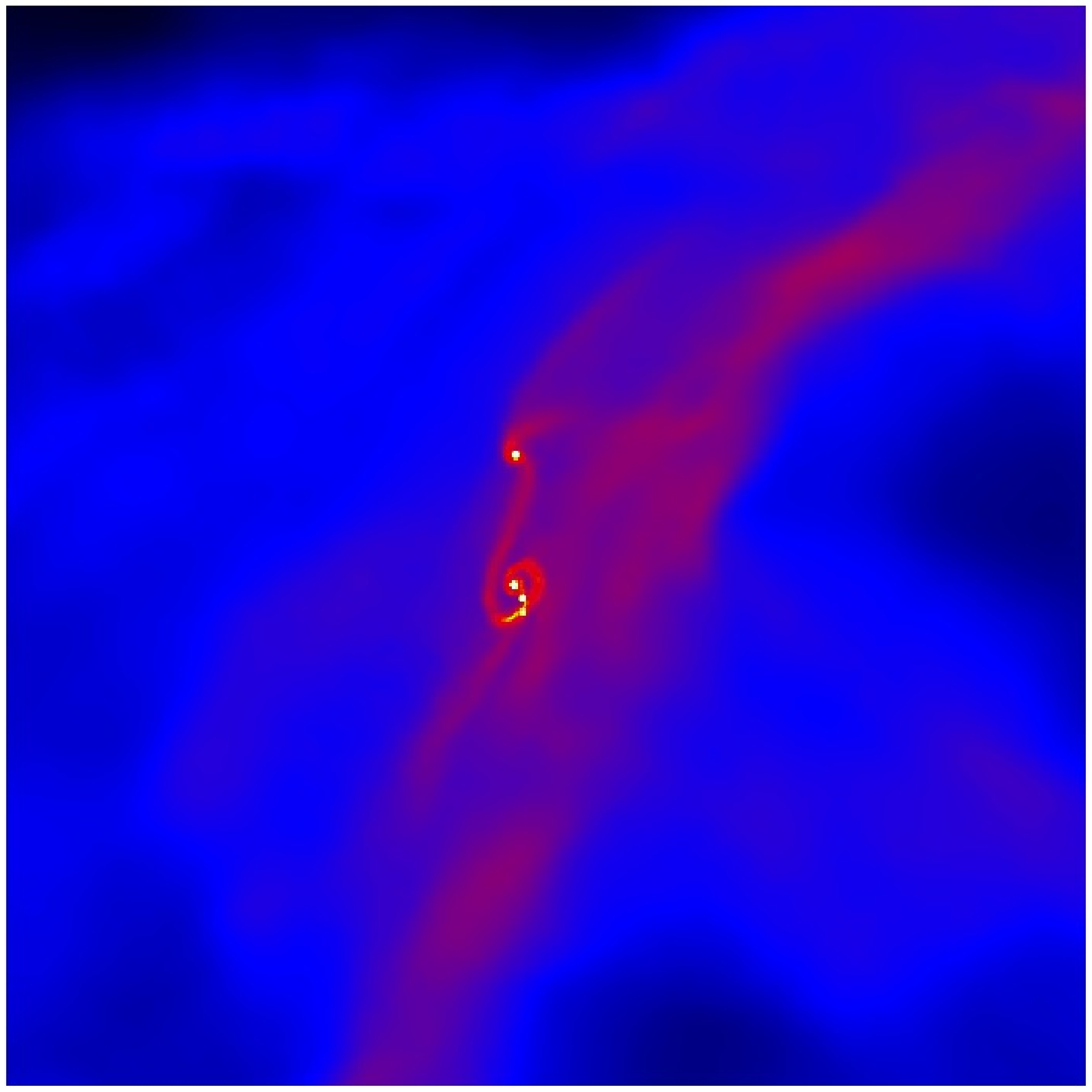,height=2.8in}  \\
\psfig{file=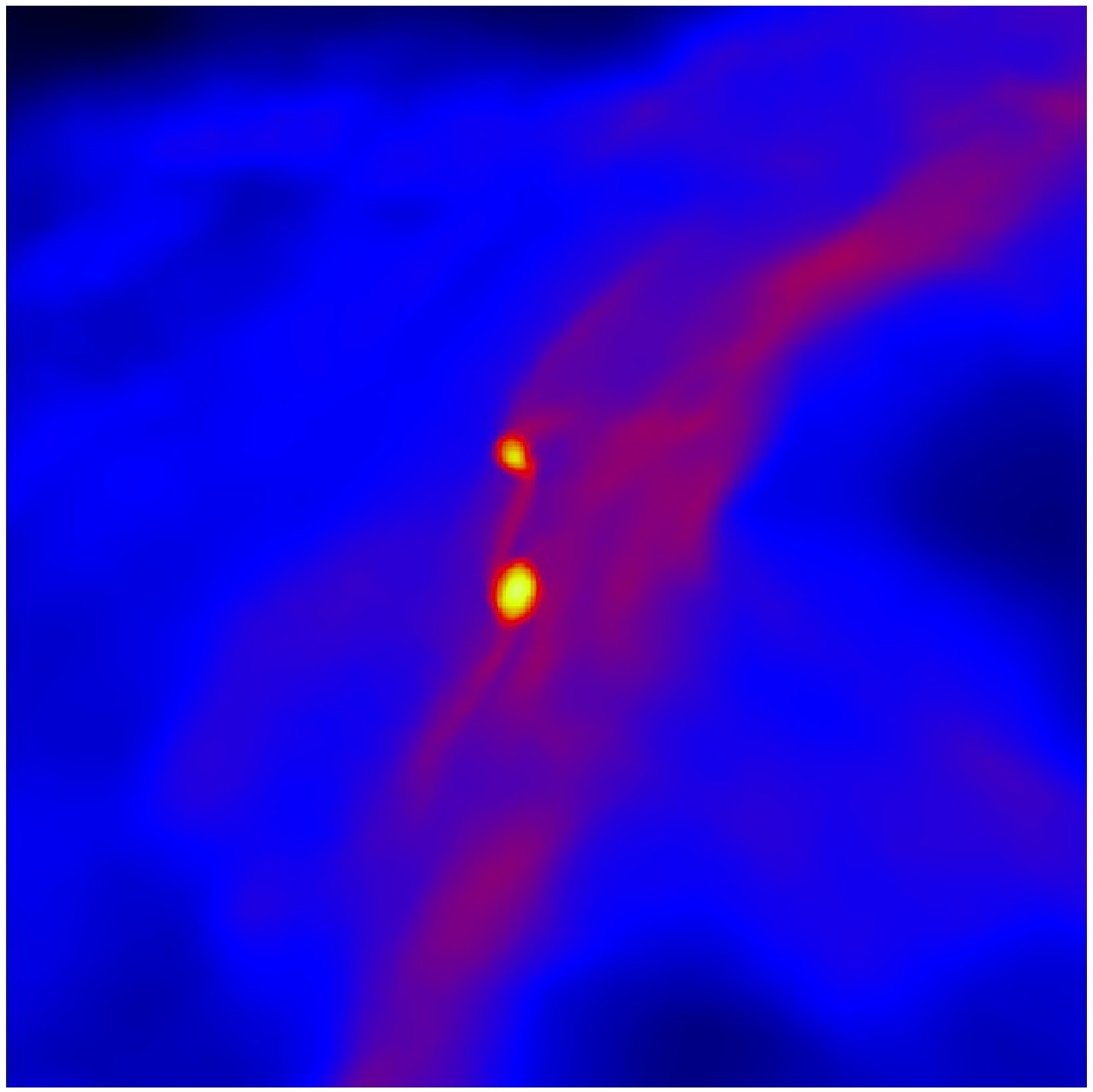,height=2.8in}  
\end{tabular}
\caption{Logarithmic column density plots
(as Figure~1, minimum and maximum densities of 
$4 \times 10^{-17}$gcm$^{-3}$ and $4 \times 10^{-13}$gcm$^{-3}$) when $\alpha=0.25$, R=0.06pc for isothermal 
(Model~1, top) and non-isothermal (Model~4, bottom) cases. The size of the plots
are 0.02pc by 0.02pc and the
time is $0.32 t_{ff}$, $t_{ff}=4.5 \times 10^4$ years.}
\end{figure}

\begin{figure*}
\centerline{\psfig{file=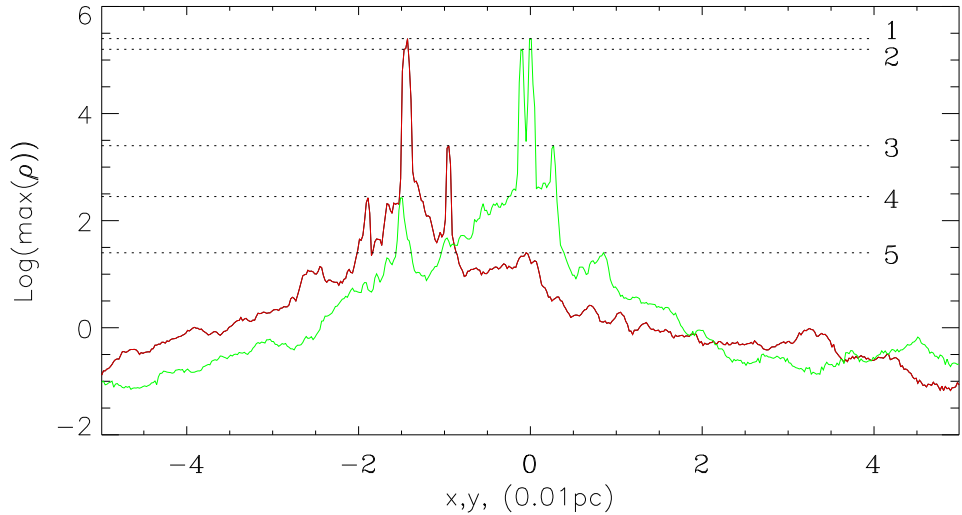,height=2.8in}}
\caption{Logarithmic plot of maximum density (over all points in the simulation) versus
x (red) and versus y (green) for Model~4. Dotted lines show 5 dominant peaks. $\rho$
measured in computational units where 1 unit=$7 \times 10^{-17}$gcm$^{-3}$. 
The plot is taken at $t=0.53t_{ff}$  ($t_{ff}=4.5 \times 10^4$ years) when the 
equation of state is non-isothermal.} 
\end{figure*}

\begin{figure}
\centering
\psfig{file=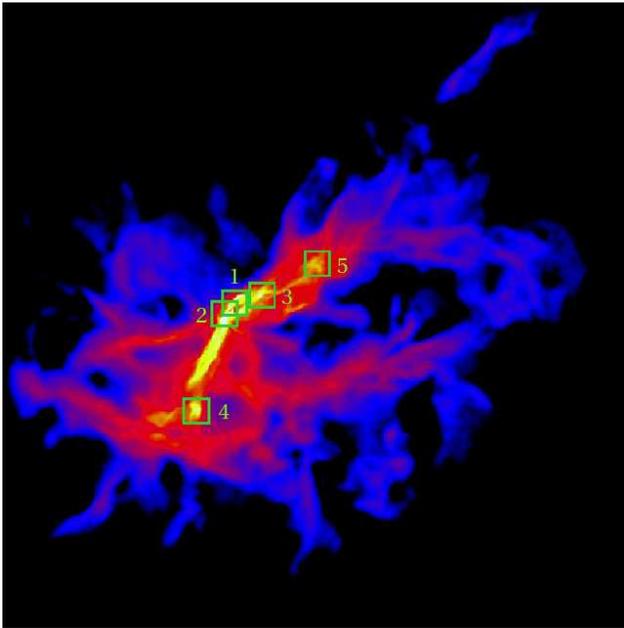,height=3.3in}
\caption{Column density plot for Model~4, with peaks identified in Figure 4 indicated. 
Only points where 
$log(\rho)>-2$ have been selected to emphasise the structure. The size scale of this plot is 
0.1pc by 0.1pc with minimum and maximum densities of 
$7 \times 10^{-19}$gcm$^{-3}$ and $4 \times 10^{-13}$gcm$^{-3}$.
Again the plot corresponds to a time of $t=0.53t_{ff}$.} 
\end{figure}
 
\section{Conclusions}
Turbulently supported clouds of typical density structure $\rho \propto r^{-1.5}$ 
have been hypothesised to be the progenitors of massive stars. Their density 
concentration could prevent fragmentation and their high accretion rates overwhelm the radiation
pressure from the accreting massive star \citep{McKee2003,McKee2002}.
We have performed numerical simulations of this scenario and
found that the initial centrally condensed density profile proved insufficient to prevent
fragmentation. Turbulent support generates significant
structure in the core which forms the basis for subsequent fragmentation. 
Turbulence
cannot be assumed to act as the equivalent of an isotropic pressure.
In addition, although the total mass accretion rates are comparable with MT(2003), 
individual protostellar accretion rates are significantly lower.

The fragmentation of the centrally condensed $30M_{\odot}$ cores forms $\approx 20$
stars over the time period of our isothermal simulations. The protostars
form from a combination of the fragmentation of filamentary structure and the
fragmentation of rotationally supported discs. Our non-isothermal runs suppress
this disc fragmentation as the gas is assumed to heat on scales of the discs.
Nevertheless, additional fragmentation is expected once a second collapse
phase softens the equation of state. Competitive accretion in these clusters
will determine the eventual stellar masses \citep{BCBP2001}.

These simulations neglect the potential effects of any magnetic fields
present in the core. We do note that this should not impede the fragmentation
process as magnetic fields have been shown not to affect the structure
generation in turbulent molecular clouds \citep{Stone1998,Padoan2004}. Even
in the presence of magnetic fields, turbulence does not act as an isotropic
support, and neither turbulence nor magnetic fields should be modelled as an 
isotropic pressure term. 
    
\bibliographystyle{mn2e}
\bibliography{test}

\bsp

\label{lastpage}

\end{document}